\documentclass[lettersize,journal]{IEEEtran}
\usepackage{cite}
\usepackage{amsmath,amssymb,amsfonts}
\usepackage{algorithmic}
\usepackage{graphicx}
\usepackage{textcomp}
\usepackage{xcolor}
\def\BibTeX{{\rm B\kern-.05em{\sc i\kern-.025em b}\kern-.08em
    T\kern-.1667em\lower.7ex\hbox{E}\kern-.125emX}}

\usepackage{url}
\usepackage[T1]{fontenc}
\usepackage[latin1]{inputenc}
\usepackage{soulutf8}
\usepackage{tikz}
\usepackage{pgfplots}
\usepackage{bm}
\usepackage{cite}
\usepackage{acro} \acsetup{format/foreign = {\emph}}
\usepackage{cleveref}

\newtheorem{definition}{Definition}

\DeclareMathOperator{\sinc}{sinc}
\DeclareAcronym{FDM}{
	short = FDM,
	long  = frequency-division multiplexing,
	tag = acrs,
}
\DeclareAcronym{OFDM}{
	short = OFDM,
	long  = orthogonal frequency-division multiplexing,
	tag = acrs,
}

\DeclareAcronym{CFR}{
	short = CFR,
	long  = channel frequency response,
	tag = acrs,
}

\DeclareAcronym{CSI}{
	short = CSI,
	long  = channel state information,
	tag = acrs,
}

\DeclareAcronym{OFDMA}{
	short = OFDMA,
	long  = orthogonal frequency-division multiple access,
	tag = acrs,
}

\DeclareAcronym{O-OFDM}{
	short = O-OFDM,
	long  = optical orthogonal frequency division multiplexing,
	tag = acrs,
}

\DeclareAcronym{CD}{
	short = CD,
	long  = chromatic dispersion,
	tag = acrs,
}

\DeclareAcronym{PMD}{
	short = PMD,
	long  = polarization mode dispersion,
	tag = acrs,
}

\DeclareAcronym{3-D}{
	short = 3-D,
	long  = three-dimensional,
	tag = acrs,
}

\DeclareAcronym{2-D}{
	short = 2-D,
	long  = two-dimensional,
	tag = acrs,
}

\DeclareAcronym{4-D}{
	short = 4-D,
	long  = four-dimensional,
	tag = acrs,
}

\DeclareAcronym{ELU}{
	short = ELU,
	long  = exponential linear unit,
	tag = acrs,
}

\DeclareAcronym{GELU}{
	short = GELU,
	long  = Gaussian error linear unit,
	tag = acrs,
}

\DeclareAcronym{RELU}{
	short = RELU,
	long  = rectified linear unit,
	tag = acrs,
}
\DeclareAcronym{SELU}{
	short = SELU,
	long  = scales exponential linear unit,
	tag = acrs,
}
\DeclareAcronym{LTE}{
	short = LTE,
	long  = long term evolution,
	tag = acrs,
}
\DeclareAcronym{4G}{
	short = 4G,
	long  = fourth generation,
	tag = acrs,
}
\DeclareAcronym{SDMA}{
	short = SDMA,
	long  = space division multiple access,
	tag = acrs,
}
\DeclareAcronym{SDMA-OFDM}{
	short = SDMA-OFDM,
	long  = space division multiple access orthogonal frequency-division multiplexing,
	tag = acrs,
}
\DeclareAcronym{NOMA}{
	short = NOMA,
	long  = non-orthogonal multiple access,
	tag = acrs,
}
\DeclareAcronym{MIMO}{
	short = MIMO,
	long  = multiple-input-multiple-output,
	tag = acrs,
}
\DeclareAcronym{MIMO-NOMA}{
	short = MIMO-NOMA,
	long  = multiple-input-multiple-output non-orthogonal multiple access,
	tag = acrs,
}
\DeclareAcronym{NN}{
	short = NN,
	long  = neural network,
	tag = acrs,
}

\DeclareAcronym{LOS}{
	short = LOS,
	long  = line-of-sight,
	tag = acrs,
}
\DeclareAcronym{NLOS}{
	short = NLOS,
	long  = non-line-of-sight,
	tag = acrs,
}
\DeclareAcronym{WINNER}{
	short = WINNER,
	long  = Wireless World Initiative for New Radio,
	tag = acrs,
}
\DeclareAcronym{QuaDRiGa}{
	short = QuaDRiGa,
	long  = quasi deterministic radio channel generator,
	tag = acrs,
}
\DeclareAcronym{MT}{
	short = MT,
	long  = mobile terminal,
	tag = acrs,
}
\DeclareAcronym{BS}{
	short = BS,
	long  = base station,
	tag = acrs,
}
\DeclareAcronym{NR}{
	short = NR,
	long  = new radio,
	tag = acrs,
}
\DeclareAcronym{AWGN}{
	short = AWGN,
	long  = additive white Gaussian noise,
	tag = acrs,
}
\DeclareAcronym{SNR}{
	short = SNR,
	long  = signal-to-noise ratio,
	tag = acrs,
}
\DeclareAcronym{ADAM}{
	short = ADAM,
	long  = adaptive moment estimation,
	tag = acrs,
}
\DeclareAcronym{MSE}{
	short = MSE,
	long  = mean squared error,
	tag = acrs,
}
\DeclareAcronym{ROC}{
	short = ROC,
	long  = receiver operating characteristic,
	tag = acrs,
}

\DeclareAcronym{ML}{
	short = ML,
	long  = machine learning,
	tag = acrs,
}

\DeclareAcronym{5G}{
    short = 5G,
    long  = fifth generation,
    tag = acrs,
}

\DeclareAcronym{6G}{
	short = 6G,
    long  = sixth generation,
	tag = acrs,
}

\DeclareAcronym{TAS}{
	short = TAS,
    long  = transmit antenna selection,
	tag = acrs,
}

\DeclareAcronym{DL}{
	short = DL,
    long  = deep learning,
	tag = acrs,
}

\DeclareAcronym{CNN}{
	short = CNN,
    long  = convolutional neural network,
	tag = acrs,
}
\graphicspath{{Figures/}}
\pgfplotsset{compat=1.17}
\usepackage{siunitx}

\begin{document}

\title{An Efficient Machine Learning-based Channel Prediction Technique for OFDM Sub-Bands}

\author {Pedro E. G. Silva, \IEEEmembership{Graduate Student Member,~IEEE}, Jules~M.~Moualeu, \IEEEmembership{Senior Member,~IEEE},\\  Pedro H. Nardelli, \IEEEmembership{Senior Member,~IEEE}, and Rausley A. A. de Souza, \IEEEmembership{Senior Member,~IEEE}
%\thanks{This work is carried out under BRICS Multilateral R\&D Project with No. DST/IMRCD/BRICS/PILOTCALL2/LargEWiN/2018(G). The research of P. K. Upadhyay is supported in part by the project under the Visvesvaraya PhD Scheme of Ministry of Electronics \& Information Technology (MeitY), Government of India, being implemented by Digital India Corporation (formerly Media Lab Asia).}
\thanks{This work is supported in part by Academy of Finland via: (a) FIREMAN consortium n.326270 as part of CHIST-ERA-17-BDSI-003,  (b) EnergyNet Fellowship n.321265/n.328869/n.352654, and (c) X-SDEN project n.349965; by Jane and Aatos Erkko Foundation via STREAM project; by  CNPq (Grant  311470/2021-1), by S\~{a}o Paulo Research Foundation (FAPESP) (Grant No. 2021/06946-0), by CAPES (Grant No. 88887.353680/2019-00), and by RNP, with resources from MCTIC, Grant No. 01245.010604/2020-14, under the Brazil 6G project of the Radiocommunication Reference Center of the National Institute of Telecommunications, Brazil. This work is also ssupported in part by the South African National Research Foundation (NRF) under the BRICS Multilateral Research and Development Project (Grant No. 116018).}
\thanks{P. E. G. Silva and P. H. Nardelli are with the School of Energy Systems, Lappeenranta-Lahti University of Technology, Lappeenranta, Finland (e-mails: \{pedro.goria.silva, pedro.nardelli\}@lut.fi).}
\thanks{J. M. Moualeu is with the School of Electrical and Information Engineering, University of the Witwatersrand, Johannesburg 2000, South Africa (e-mail: jules.moualeu@wits.ac.za).}
\thanks{R. A. A. de Souza is with the National Institute of Telecommunications (Inatel), Santa Rita do Sapuca\'{i}, Brazil (e-mail: rausley@inatel.br).}
}

\maketitle
\begin{abstract}
The acquisition of accurate \ac{CSI} is of utmost importance since it provides performance improvement of wireless communication systems. 
However, acquiring accurate \ac{CSI}, which can be done through channel estimation or channel prediction, is an intricate task due to the complexity of the time-varying and frequency selectivity of the wireless environment. 
%
% Traditional approaches used to estimate or predict channel information may not be accurate as they generally rely on unrealistic assumptions. 
%
To this end, we propose an efficient \ac{ML}-based technique for channel prediction in \ac{OFDM} sub-bands. The novelty of the proposed approach lies in the training of channel fading samples used to estimate future channel behaviour in selective fading.

\end{abstract}

%\acresetall

\section{Introduction} \label{sec: Intro}

% With the advances in communication technology and multimedia services, the demand for high bandwidth is becoming increasingly imperative \textcolor{red}{[R1]}-\textcolor{red}{[R3]}. 
% 
\IEEEPARstart {T}he increase in user density and the improvement of spectral efficiency requirements for various wireless communication systems have driven the development of multi-carrier modulation techniques. 
For example, \acf{OFDM} has been implemented in the current \ac{5G} networks and have emerged as a promising candidate for future \ac{6G} communication systems thanks to its robustness against multipath fading and its great performance in terms of spectral efficiency~\cite{LiBaolong2021LOOW}. %LiBaolong2022RHOO,
One important aspect to ensure reliable transmission in \ac{OFDM} systems is the acquisition of accurate \acf{CSI}. However, estimating channel information in a precise manner is challenging and in some cases impractical due to the complexity of the time-varying wireless environment. 
Consequently, the channel coefficients may be outdated during the estimation process as the channels may have already changed due to its rapid fluctuations within two consecutive packets \cite{jmoualeu}, and thus lead to the performance degradation of the system.
To overcome this issue, channel prediction can be exploited. Unlike channel estimation, channel prediction allows to forecast the estimated channel information for future channel responses \cite{duel-hallen}.
Furthermore, channel prediction reduces the transmitted overhead information and therefore, does not negatively impact the date rate.

A large number of sub-carriers are employed in \ac{OFDM} systems to combat the deleterious effects of frequency selective fading. 
Moreover, the use of \ac{OFDM} over frequency-selective fading channels can be advantageous due to the breakdown of the available frequency band into sub-bands. 
Since the coherence bandwidth is narrow compared to the available bandwidth, it is possible to slice the spectrum into sub-bands of uncorrelated fading. 
Consequently, the effects of frequency selective fading may only degrade one sub-band of a multi-carrier system. 
However, such degradation could be severe on the system performance and could be avoided through efficient sub-band allocation or channel prediction. 
Some studies have proposed techniques to address: (a) the sub-band allocation problem in \ac{4G} mobile networks, such as \ac{SDMA-OFDM}~\cite{Vandenameele895036}; (b) the channel prediction issue in adaptive \ac{OFDM} systems~\cite{icwong}. %Alansi5876975,BAGADI201693
However, traditional approaches used for channel prediction (e.g., in reference \cite{icwong}) consist of a tedious estimation process of propagation parameters and often rely on impractical assumptions yielding inaccurate \ac{CSI}. 

Deep learning and \acp{NN} are excellent tools that can be used for event prediction and pattern Recognition~\cite{Torres2021}. %Sornam8441512  
Recently, \ac{ML} has been applied in \ac{OFDM}-based systems to improve the accuracy of channel prediction techniques~\cite{wjiang,liulei,yshao}. In \cite{wjiang}, a prediction-enabled \ac{MIMO}-\ac{OFDM} with \ac{TAS} is investigated, while the work of \cite{liulei} proposes a deep \ac{NN}-based approach for channel prediction in an underwater acoustic \ac{OFDMA} system. In \cite{yshao}, a \ac{DL}-based channel prediction is proposed in an effort to reduce the demand of pilot symbols in \ac{OFDM} sytems.
Different from \cite{wjiang,liulei,yshao}, our proposed channel prediction approach based on \ac{CNN} assumes that the fading behaviour of a particular channel exhibits some peculiar characteristic, i.e., a signature that represents how physical objects are acting within the propagation environment.
However, a mathematical model capable of accurately predicting all possible multipaths that the transmitted signal can travel through to the receiver is not trivial and is generally intractable. 
Motivated by the preceding discussion, we aim to address this challenge by adopting a \ac{ML} approach that efficiently predicts the fading behaviour of a channel based on past-trained samples. 
Hence, we focus on the forecast of unsuitable sub-bands used for signal transmission.
Furthermore, we obtain some promising results with the aid of a \ac{CNN} method operating in the time-frequency domain. 

The proposed technique can find some benefits in \ac{6G} use-case scenarios; for instance, low-latency applications would be able to mitigate the delay due to retransmissions by properly predicting the state of the channel at a future point and thus selecting the most appropriate time and sub-band for the transmission.
Furthermore, this work can be beneficial to several proposed use cases for \ac{5G}, especially environments where there is an equipment powered by batteries that may benefit from a prediction of the fading behaviour by (i) saving energy with unmodulated carrier transmission for channel estimation and (ii) transmitting in bands with less severe fading.
% , such as a smart grid, Energy Internet. 
% 
These environments may require a high density of equipment and extremely efficient use of the battery.

The remainder of the paper is organised as follows. 
In \Cref{sec: Channel}, the method adopted for the simulation of the channel is described and a brief discussion about other mobile radio channel sample generation techniques is carried out.
We introduce and discuss the adopted \ac{CNN} for channel prediction in \Cref{sec: Chan_Pred}.
We assess the performance and some remarks of the proposed predictor through numerical results in \Cref{sec: Num Results}.  
The work ends with brief conclusions in \Cref{sec: Conc}.

% 
% The remainder of this work is structured as follows:
% 
%\vspace{-0.125in}
\section{Channel Model Simulator} \label{sec: Channel}

Although the effects of an antenna (mainly its radiation pattern) are relevant in the description of the received signal, we can omit them---at least for now and consider their analysis in future works---from our channel simulations.
Namely, modern mobile radio channel simulators deal separately with relevant aspects of wave propagation and antenna characteristics, such as \ac{WINNER} II~\cite{bultitude20074} and \ac{QuaDRiGa}~\cite{Jaeckel6758357}. 
Generally speaking, the radiation pattern of the receiving antenna only weighs in the received signal envelope, but not the behaviour of the multipath clusters over time~\cite{Yin7378822}.
Hence, we omit the impact of antenna parameters on the channel response\footnote{However, it will be unquestionably of great value to include the effect of the radiation pattern of the antennas in future studies.}. 
Another simplification adopted is the evaluation of the channel on a \ac{2-D} plane through which the mobile radio channel simulator consists of reflection points of the signal transmitted in the \ac{2-D} plane.
Keeping that in mind, the simplified model which allows some reflecting points to move randomly during the channel simulation can be used to address the birth and death problem of multipath clusters.

Here, our goal is to confirm that the frequency response of the mobile radio channel may exhibit predictable behaviour patterns over the following conditions: 
(i) a \ac{BS} covers a wide area; 
(ii) \ac{BS} uses an omnidirectional antenna; 
(iii) a single set of large-scale parameters can effectively describe the simulated environment; 
(iv) the simulated environment is such that there are multipaths.

In what follows, we focus our discussion on this random movement of the reflection points. 
Suppose that the underlying environment is a densely populated area (i.e., city centre) with cars, pedestrians, buildings, etc, that could reflect or spread the signal. 
We aim to predict the channel behaviour over ten frames of the \ac{5G} \ac{NR}. 
Consequently, our proposed approach predicts the channel behaviour up to $100$ ms in the future since each \ac{5G} \ac{NR} frame has a period of $10$ ms~\cite{TSGS2019}. 
Given the conditions of the movement and acceleration inherent to these reflective objects, the $100$ ms is undoubtedly a despicable time to expect any dramatic change. 
In other words, given a relatively short time, we can expect that the reflection points will either provide short linear routes with constant speed or remain static. 
Thus, it can be assumed that the reflection points move at a constant speed and direction over time.

We now present a step-by-step guide for generating the channel samples. To begin with, an environment layout to be created should contain the following pieces of information: the relative position of the transmitter $\boldsymbol{P_\text{tx}}$ in Cartesian coordinates, the relative position of the receiver $\boldsymbol{P_\text{rx}}$ in Cartesian coordinates, the speed vector of the receiver $\boldsymbol{S_\text{rx}}$, the total number of reflection points $N_\text{r}$, the number of mobile reflection points $N_\text{m}$, the initial position of the reflection points $\boldsymbol{P_\text{r}}$ (randomly generated), the speed vector of the reflection points $\boldsymbol{S}$, the carrier frequency $f_\text{c}$, the sampling frequency $f_\text{s}$, the bandwidth of the transmitted signal $B$, and the sampling time window $p$. 
Notice that $\boldsymbol{P_\text{r}}$, $\boldsymbol{S_\text{rx}}$, and $\boldsymbol{S}$ have their coordinates generated according to a Gaussian distribution $\mathcal{N}(\mu_\text{p}, \sigma_\text{p}^2)$, $\mathcal{N}(\mu_\text{rx}, \sigma_\text{rx}^2)$ and $\mathcal{N}(\mu_\text{s}, \sigma_\text{s}^2)$, respectively, and only $N_\text{m}$ positions of the vector $\boldsymbol{S}$ are not null.

Moreover, we calculate the channel impulse response by first obtaining the expression of the length of a multipath $l_i$, for $i=\{1,2,\dots,N\}$, through the Euclidean norm 
%\begin{equation}
    $l_i=|\boldsymbol{P}_i-\boldsymbol{P_\text{rx}}|+|\boldsymbol{P}_i-\boldsymbol{P_\text{tx}}|$, 
%\end{equation}
where $\boldsymbol{P_\text{r}}=[\boldsymbol{P}_1, \boldsymbol{P}_2, \dots, \boldsymbol{P}_N]$. The delay and phase of the multipath $i$ can be calculated using $\tau_i=l_i/c$ and $\phi_i=-l_i f_\text{c}/c \mod{2 \pi}$, respectively, in which $c$ represents the speed of light. Considering the effect of the free space attenuation for each multipath, the channel impulse response at time $t$ in the baseband is given by
\begin{equation}
    I=\sum^{N}_{i=1} \frac{c}{4 \pi f_\text{c} l_i} \exp{(\phi_i + 2\pi D_i l_i)} \delta(\tau_i - t),
\end{equation}
where $\delta(\cdot)$ is the Dirac delta function and $D_i$ is the Doppler shift. For simplicity, we adjust the delays $\tau_i$ in such a way that the shortest route has a delay equal to zero.
% the convolution of the transmitted signal with the impulse response channel.

% Knowing that transmitted reference signal has a limited duration of $p$ seconds, that is, it is truncated in time, assuming a flat frequency for transmitted signal (i.e., $s_t(t)=\sinc(B\,t)$), and sampling the received signal $s_r$ at time $k = \mathcal{I}/f_\text{s}$ with $\mathcal{I}={...,-1,0,1,...}$, we have
It is known that the receiver only has a short $p$-second window to estimate the channel, assuming that the transmitted signal is given by
% that the transmitted signal undergoes flat-frequency fading  (i.e., 
$s_\text{tx}(t)=\sinc(B t)$, and sampling the received signal $s_\text{rx}(t)$ at a rate of $f_\text{s}$, we have 
%\begin{equation}
    $s_\text{rx}(k)= \sinc(B \, k) \ast I+n$, $0\le k \le p$, 
%\end{equation}
in which $n$ denotes the \ac{AWGN} and $\ast$ stands for the convolution operation.

The frequency response of the channel is given by the discrete Fourier transform $\mathcal{F}(\cdot)$ of the samples collected from the received signal, and its modulus can be evaluated by 
%\begin{equation}
    $C(f)=|\mathcal{F}(s(k))|$. 
%\end{equation}
After obtaining the frequency response of the channel, the positions of the receiver and other mobile points are updated as
\begin{align}
    \boldsymbol{P_\text{rx}} \leftarrow \boldsymbol{P_\text{rx}}+ \boldsymbol{S_\text{rx}} \Delta t  \\ 
    \boldsymbol{P_\text{r}} \leftarrow \boldsymbol{P_\text{r}}+ \boldsymbol{S} \Delta t,
\end{align}
% and
% \begin{align}
%     \boldsymbol{P_\text{r}} \leftarrow \boldsymbol{P_\text{r}}+ \boldsymbol{S} \Delta t
% \end{align}
where $\Delta t$ stands for the time between two consecutive channel simulations. Once the updates of these positions are complete, the steps previously taken to obtain the channel frequency response are repeated. 

\begin{figure}%[t] %ht!
 	\centering
 	\includegraphics[width=1\linewidth]{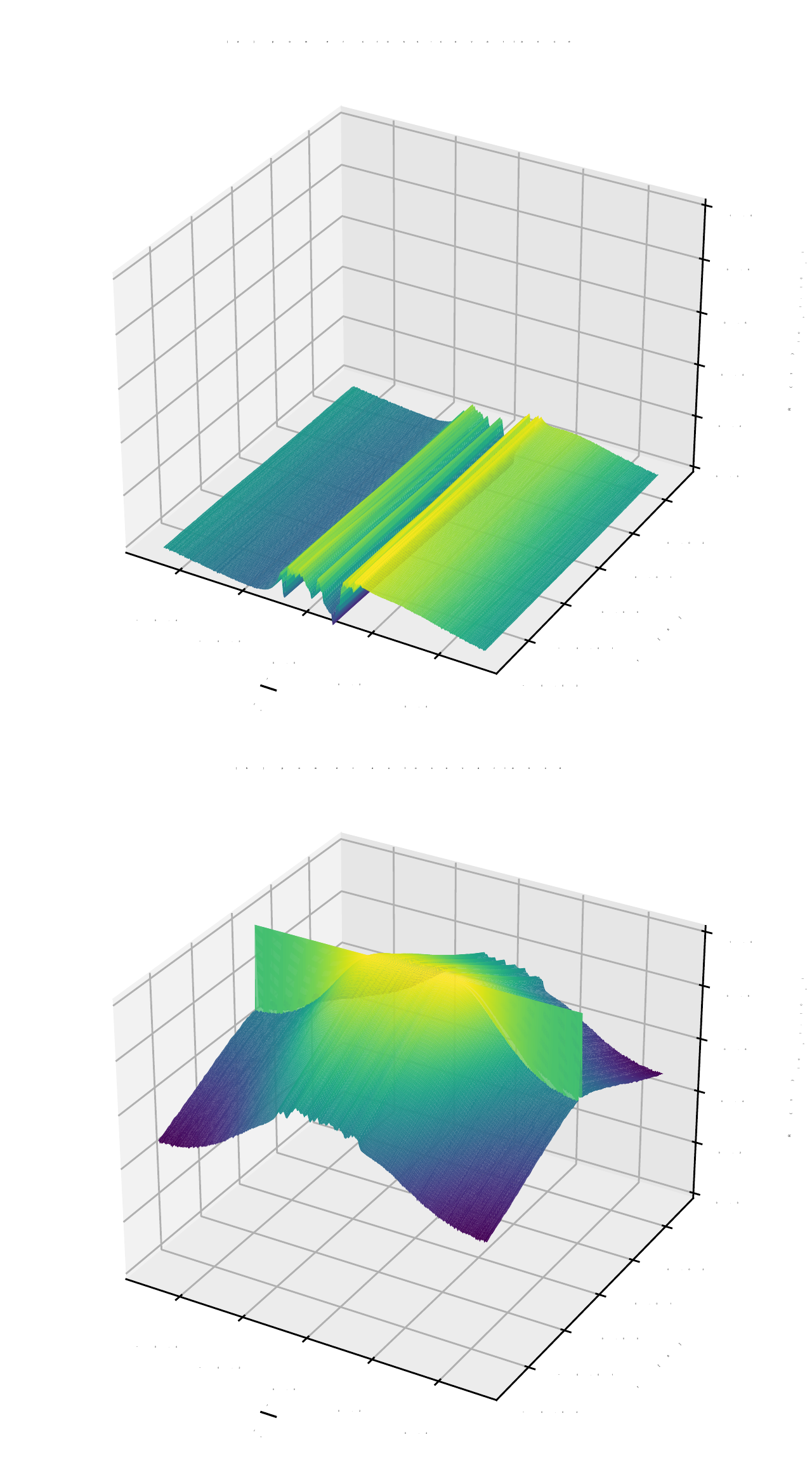}
 	\caption{(a) Cross-covariance and (b) autocovariance over $\tau$ for the frequency band of interest.}
	\label{fig: 3D Time x Frequency}
 \end{figure}

\begin{definition}[Covariance]
    Let $x_k$ and $y_k$ be the $k$-th sample of stochastic processes $X(t)$ and $Y(t)$, respectively. 
    We then define the normalised covariance function by 
    \begin{equation}
        R_{X,Y}(\tau)= \frac{1}{\sqrt{\text{VAR}(x)\text{VAR}(y)}}\sum_j x_j y_{j-\tau},
    \end{equation}
    where $\text{VAR}(\cdot)$ represents the variance of the samples. 
\end{definition}
\begin{figure}%[hb!]
 	\centering
\includegraphics[width=1.2\linewidth]{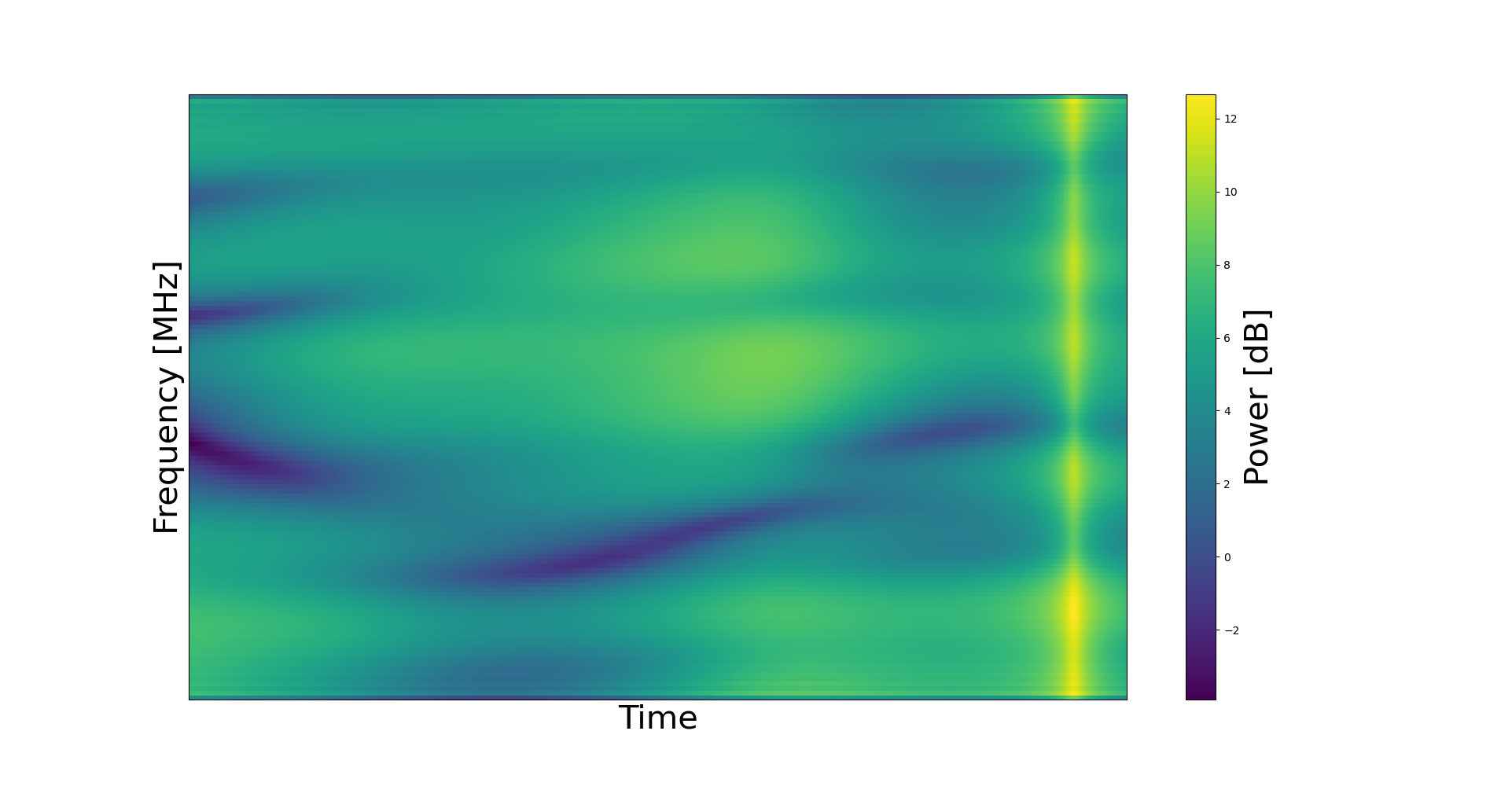}
 	\caption{Frequency response of the channel over time.}
	\label{fig: 2D Time x Frequency}
 \end{figure}
Figure \ref{fig: 2D Time x Frequency} depicts the frequency response over time for one simulation of the mobile radio channel in the frequency bands as the metric of interest. 
Moreover, Fig. \ref{fig: 3D Time x Frequency} (a) shows cross-covariance $R_{|C_0|,|C_1|}(\tau)$ between the magnitude (power) of the frequency response $C_0(f)$ and $C_1(f)$ of two distinct channels (i.e., two completely different initial conditions of channel simulation).
The following simulation parameters are used: $f_\text{c}=900$ MHz, $f_\text{s}=51.2$ MHz, $B=12.8$ MHz, $p=10$ $\mu$s, $\mu_\text{p}=0$, $\sigma^2_\text{p}=4900$, $\mu_\text{rx}=1$, $\sigma^2_\text{rx}=4$, $\mu_\text{s}=0$, $\sigma^2_\text{s}=30.25$ and 12 dB of \ac{SNR}.
In addition to these parameters, we have $N_\text{r}=100$ and $N_\text{m}=63$ for Fig.~\ref{fig: 2D Time x Frequency}, $N_\text{r}=300$ and $N_\text{m}=200$ for $C_0(f)$ in Fig.~\ref{fig: 3D Time x Frequency}, and $N_\text{r}=2$ and $N_\text{m}=2$ for $C_1(f)$ in Fig.~\ref{fig: 3D Time x Frequency} (a).
The presence of deep fading at specific frequencies---in some cases, corrupting an entire sub-band---is evident in Fig. \ref{fig: 2D Time x Frequency}.
It is also clear that our simulation of the channel behaves dynamically over time. 
One can notice in Fig. \ref{fig: 3D Time x Frequency} that (i) the proposed channel simulation is capable of generating two uncorrelated channels and (ii) guaranteeing a temporal covariance that decreases rapidly with increasing $\tau$. 
Therefore, at least for the purpose of this work, we can say that the simulation of the channel was effective.

%\vspace{-0.165in}
\section{Channel Prediction} \label{sec: Chan_Pred}
% \hl{One and half page is expected!}
% 
To obtain an accurate measure of the channel transfer function is a challenging task and this appears to be intractable from a practical viewpoint. To circumvent this issue, a widespread technique with relatively good accuracy that consists of transmitting a reference signal embedded in the transmitted signal can be adopted. Consequently, the estimate of the channel transfer function from the received signal becomes feasible. Up to this point, the time dependence of the \ac{CFR} has been conveniently omitted; nevertheless, it can be directly associated as follows. 
Let $F$ be the total number of sub-carriers for which channel state information is available, $D$ be the total number of steps ahead desired for simultaneous forecasting (whole prediction span), and $T$ be the time length of the input layer of the proposed \ac{NN} (number of consecutive observations of the \ac{CFR} over time). In addition, we denote $\bm{H}_j \in \mathbb{C}^{F}$ as the sequence of complex samples of the \ac{CFR} in the $j$th estimation. Therefore, we have a sampling of the response channel at time $t = j \Delta t$ seconds. Due to the \ac{AWGN} and the temporal truncation of the received signal, $\bm{H}_j$ is a noisy and slightly distorted version of the \ac{CFR}. 

Since our interest remains in feeding the \ac{NN} with the channel samples, it is crucial to adjust the vector $\bm{H}_j$. Thus, one needs to first convert the observation series $\bm{H}_j$ into \ac{3-D} tensors $X \in \mathbb{C}^{(J \times F \times T)}$, in which the index $j$ represents the time length of the tensor, i.e., the sample number of the \ac{CFR}. Then, as the proposed \ac{NN} requires a real sample set for its entry, another dimension to the tensor $X$ needs to be added. This transformation can mathematically be expressed \begin{subequations}
\begin{equation}
    X_{j,f,i,0} = \operatorname{Re}(H_{j-T+i}(f)), \end{equation}
    \begin{equation}X_{j,f,i,1} = \operatorname{Im}(H_{j-T+i}(f)), \end{equation}
    \label{eq: transf Input}
\end{subequations}
$\forall$ $j,f$ and $i$. Now, the tensor $X$ with dimensions $(j\times H \times T \times 2)$ can be used as an input of the underlying \ac{NN}.
 
The prediction problem consists of optimising the mathematical model parameters, $\Gamma$, called predictor, such that the proposed prediction is as close as possible to the training labels. For any discrete-time $j$, the training labels are $\bm{H}_{ \{j+1,j+2,\dots,j+D\} }$, and the tensor of input variables for the \ac{NN} is given by $X_{j,f,i,c}$, $\forall$ $f,i$ and $c$. The Gamma predictor $\Gamma(\cdot,\zeta)$ is parameterised by a real-valued vector of parameters $\boldsymbol{\zeta} \in \mathbb{R}^{Z}$.
The output of the underlying \ac{NN} is also a \ac{4-D} tensor $y \in \mathbb{R}^{(j \times F \times T \times 2)}$. Given an input tensor $X$, one can evaluate $y$ with the predictor $\Gamma$ as $y = \Gamma(X,\boldsymbol{\zeta})$. In the sequel, we deal with some aspects pertinent to the prediction model $\Gamma$, and we also describe the transformation process that occurs internally in the proposed \ac{NN}.

% 
%\vspace{-0.20in}
\subsection{Causality}
Causality is a condition inherent in any physical model of temporal prediction. 
Since the model must estimate future outcomes given past events, no mathematical operation internal to the model can be non-causal. In other words, assuming that the predictor---more precisely, one or more layers of the predictor model---has access to future events is implausible. In order to guarantee the causality of the convolution operation, temporal indices of convolution kernels consist purely of non-negative integers, and the frequency axis is symmetrical.  
% 
% \subsection{Convolutional Layers}
%\vspace{-0.20in}
\subsection{CNN Predictor} \label{subSec:Predictor}
The predictor proposed in this work is composed primarily of layers of spatial convolution, which operate in a $d$-dimensional space. 
% More details about the adopted model are presented in Section \ref{subSec:Predictor}. Herewith, the minutiae of the process and the mathematical support of a single layer of $d$-dimensional convolution are addressed. 
To simplify the notation, we denote $x \in \mathbb{R}^{(v_0 \times v_1 \times \cdots \times v_{d-1})}$ to be a $d$-dimensional tensor, in which a specific position can be accessed by $x_{u_1,\dots,u_{d-1}}$. Its indices $u_1,\dots,u_{d-1}$ will be denoted as a vector $\bm{u}$, and hereafter, $x_{u_1,\dots,u_{d-1}}$ will be referred to as $x_{\bm{u}}$.
Let $x$ and $k$ be a pair of $d$-dimensional tensors, with $d \in \mathbb{N}^{*}$. The $d$-dimensional convolution of $x$ with $k$, indicated by $x*k$, is established  as
\begin{align}
    (x*k)_{\bm{u}} := \sum_{\bm{v}}x_{\bm{u}-\bm{v}}k_{\bm{v}},
    \label{eq: conv definition}
\end{align}
$\forall$ $\bm{u}$ and $\bm{v}$ cover all terms in which $x$ and $k$ are defined.
Regarding convolutional layers of \acp{NN}, the tensor $k$ refers to the tunable parameters, and it is commonly called convolution kernel.

A kernel dilation may be used to impose a fixed integer number of consecutive null values along a dimension on the kernel tensor. 
Equivalently, given the convolution defined in \eqref{eq: conv definition}, the vector $\bm{v}$ is replaced with the index sets of the form $\{0, \Delta, 2 \Delta,\dots \}$, where $\Delta \in \mathbb{N}^*$ is an additional hyper-parameter called the dilation factor. 
Notice that, it is not mandatory to adopt identical dilation factor values for all the dimensions. In addition, the number of tunable parameters does not change when employing a dilation factor.
In an effort to classify the signal quality of a reduced set of subband for the \ac{CFR} multi-steps, we propose a structure composed of partially dilated \acp{CNN} in series.
More precisely, the proposed predictor has four partially dilated \ac{2-D} convolutional layers with activation function given by $\beta(\cdot) = \tanh(\cdot)$ 
% and kernel $k \in \mathbb{R}^{(5\times 4)}$ 
and one \ac{2-D} convolutional as the output layer. 
Table \ref{tab: predictor layers} summarises the characteristics of each of the five chosen convolutional layers, and the sequence of the layers presented follows the order in which the input data are processed. 
% An artefact is used in the first four layers through padding, and this consists of evenly padding with zeros to the left/right or up/down of the input such that the output has the same height/width dimension as the former.
% In order to mitigate possible instabilities during the training process, we add an appropriately adjusted copy of the input tensor to the output for the entire first four layers of the predictor. 
% 
\begin{table}%[hb!]
\renewcommand{\arraystretch}{1.3}
\caption{Summary of the Characteristics of the Predictor Layers.}
\label{tab: predictor layers}
\centering
\begin{tabular}{|c|c|c|c|c|}
\hline
\textbf{\#} & \textbf{Channels} & \textbf{\begin{tabular}[c]{@{}c@{}}Kernel\\ size\end{tabular}} & \textbf{\begin{tabular}[c]{@{}c@{}}Dilation \\ rate\end{tabular}} & \textbf{\begin{tabular}[c]{@{}c@{}}Activation\\ function\end{tabular}} \\ \hline
1           & 2                 & $(3 \times 10)$                                                 & $(1 \times 1)$                                                    & $\tanh(\cdot)$                                                         \\ \hline
2           & 3                & $(10 \times 10)$                                                 & $(1 \times 16)$                                                    & $\tanh(\cdot)$                                                         \\ \hline
3           & 3                & $(10 \times 10)$                                                 & $(10 \times 1)$                                                   & $\tanh(\cdot)$                                                         \\ \hline
4           & 2                 & $(10 \times 3)$                                                 & $(1 \times 64)$                                                   & $\tanh(\cdot)$                                                         \\ \hline
5           & $D$               & $(1 \times 64)$                                                & $(1 \times 1)$                                                   & exponential                                                         \\ \hline
\end{tabular}
\end{table}

\subsection{CNN Classifier}
Our classification process consists of predicting whether the channel intensity will be below a threshold. 
Thus, the proposed classifier aims to previously identify which subbands will be subject to severe fading.
The proposed classifier has four partially dilated \ac{2-D} convolutional layers identical to the predictor and one \ac{2-D} convolutional as the output layer.
The output layer of the classifier has $D$ channels, kernel size of $(1 \times 64)$, no dilation rate, and a \textit{sigmoid} as an activation function.
Note that the output is a \ac{4-D} tensor $o \in \mathbb{R}^{(j \times F \times 1 \times D)} $ with $0 \leq o \leq 1$.
\section{Numerical Results and Remarks} \label{sec: Num Results}
% \hl{Two pages are expected!}
% 
%In this section, we present and discuss the numerical results. Moreover, we qualitatively evaluate the model's forecasting capacity  and make a quantitative analysis of the predictor's accuracy. In addition, we make a comparison of the training stability between the predictor presented in Section \ref{sec: Chan_Pred} with two of its derivations.
% 
\subsection{Channel}
We perform simulation runs using Python in order to obtain the samples of the \ac{CFR}. These simulations follow the methods described throughout this paper. 
As already discussed, the transformation given by \eqref{eq: transf Input} to all samples in the channel is applied. 
% 
% Therefore, the channel samples must be understood as a tensor given by \eqref{eq: transf Input}. 
% 
The set of simulation parameters for the underlying channel is shown in Table \ref{tab: channel settings}. 
We randomly start two simulation runs (a training set is composed of two distinct channels) with the parameters described in Table \ref{tab: channel settings}. 
We accumulate the first $4096$ tensors of the complex frequency response from each simulation run to build a training set. 
The test set consists of the subsequent $512$ tensor samples of the complex frequency response from each simulation run. 
Consequently, there are $8192$ samples for the training of the predictor and $1024$ samples for the test.
\begin{table}%[h]
\renewcommand{\arraystretch}{1.3}
\caption{Summary of Channel Simulation Settings.}
\label{tab: channel settings}
\centering
\begin{tabular}{|l|c|l|}
\hline
\multicolumn{1}{|c|}{\textbf{Symbol}} & \textbf{Value} & \multicolumn{1}{c|}{\textbf{Description}}                                                                                               \\ \hline
$\boldsymbol{P_\text{tx}}$            & $(-200, 0)$    & Transmitter's relative starting position                                                                                                \\ \hline
$\boldsymbol{P_\text{rx}}$            & $(200, 0)$     & Receiver's relative starting position                                                                                                   \\ \hline
$N_\text{r}$                          & 256            & Number of reflection points                                                                                                             \\ \hline
$N_\text{m}$                          & 63             & Number of mobile reflection points                                                                                                      \\ \hline
$f_\text{c}$                          & 900 MHz        & Carrier frequency                                                                                                                       \\ \hline
$f_\text{s}$                          & 51.2 MHz       & Sampling frequency                                                                                                                      \\ \hline
$B$                                   & 12.8 MHz       & Bandwidth of the transmitted signal                                                                                                     \\ \hline
$p$                                   & 10 $\mu$s      & Sampling time window                                                                                                                    \\ \hline
$ \mu_\text{p}$                       & 0              & \begin{tabular}[c]{@{}l@{}}Mean of the Gaussian distribution used\\ for the initial position of the reflection points\end{tabular}     \\ \hline
$\sigma_\text{p}^2$                   & $70^2$         & \begin{tabular}[c]{@{}l@{}}Variance of the Gaussian distribution used\\ for the initial position of the reflection points\end{tabular} \\ \hline
$ \mu_\text{rx}$                      & 1              & \begin{tabular}[c]{@{}l@{}}Mean of the Gaussian distribution used\\ for the receiver's speed\end{tabular}                              \\ \hline
$\sigma_\text{rx}^2$                  & 4              & \begin{tabular}[c]{@{}l@{}}Variance of the Gaussian distribution used\\ for the receiver's speed\end{tabular}                           \\ \hline
$ \mu_\text{s}$                       & 0              & \begin{tabular}[c]{@{}l@{}}Mean of the Gaussian distribution used\\ for the reflection points' speed.\end{tabular}                      \\ \hline
$\sigma_\text{s}^2$                   & 100            & \begin{tabular}[c]{@{}l@{}}Variance of the Gaussian distribution used\\ for the reflection points' speed\end{tabular}                   \\ \hline
SNR                                   & 12 dB          & Signal-to-Noise ratio at receiver                                                                                                       \\ \hline
$\Delta t$                            & 500 $\mu$s     & \begin{tabular}[c]{@{}l@{}}Time between two consecutive \\ simulations of the channel\end{tabular}                                      \\ \hline
\end{tabular}
\end{table}
% 
%\vspace{-0.15in}
\subsection{Training}
% The process of training a \ac{NN} consists of optimising the parameter vector $\bm{\zeta}$ of the predictor $\Gamma(\cdot,\bm{\zeta})$. The vector $\bm{\zeta}$ represents the adjustable values of all layers of our predictor, including bias.
% % 
% In supervised learning, the vector $\zeta$ is adjusted to all available labelled training examples (i.e. training inputs paired with training targets) by minimising a loss function. 
% % 
% This loss function measures the distance between the predictions and the corresponding target values. 
% % 
% It is worth highlighting that 
We used the mean squared error and binary cross entropy as the loss function for predictor and classifier, respectively, which evaluate the multi-step ahead of predictions with the total prediction span $D$ set to 10. 
More precisely, we present to the predictor and to classifier a small (one mini-batch) set of tensors---this process is widely known as batch training or batch learning---during the training phase. 
The mini-batch has a size of 64 tensors. 
% 
% Hence, the loss function must calculate the error for each input tensor and subsequently estimate the average error of the mini-batch by an average of the individual errors. 
At the end of this process, the vector parameters $\bm{\zeta}$ are adjusted by employing refined versions of the \ac{ADAM} algorithm with a learning rate of 0.003. 
Another important concept of the training phase is Epoch which is said to 
be completed when the whole set of input samples is evaluated. 
Here, the proposed model is trained for 30 epochs.
% 
% \subsection{Training}
% 
% In order to evaluate the training and convergence of the proposed model, two predictors are evaluated, i.e., Predictor 1 (Pred 1) and Predictor 2 (Pred 2). Both predictors are made up of the layers described in Table \ref{tab: predictor layers}, and both use the same training method described earlier. 
% % 
% However, due to the instability and non-convergence of Predictor 1, a technique known as \textit{skip} is adopted in Predictor 2 as follows.
% % 
% In order to mitigate possible instabilities during the training process, an appropriately adjusted copy of the input tensor to the output for all the first four layers of the Predictor 2 is added. 

\begin{figure}%[h]%t!]
    \centering
    \includegraphics[width=0.9\linewidth]{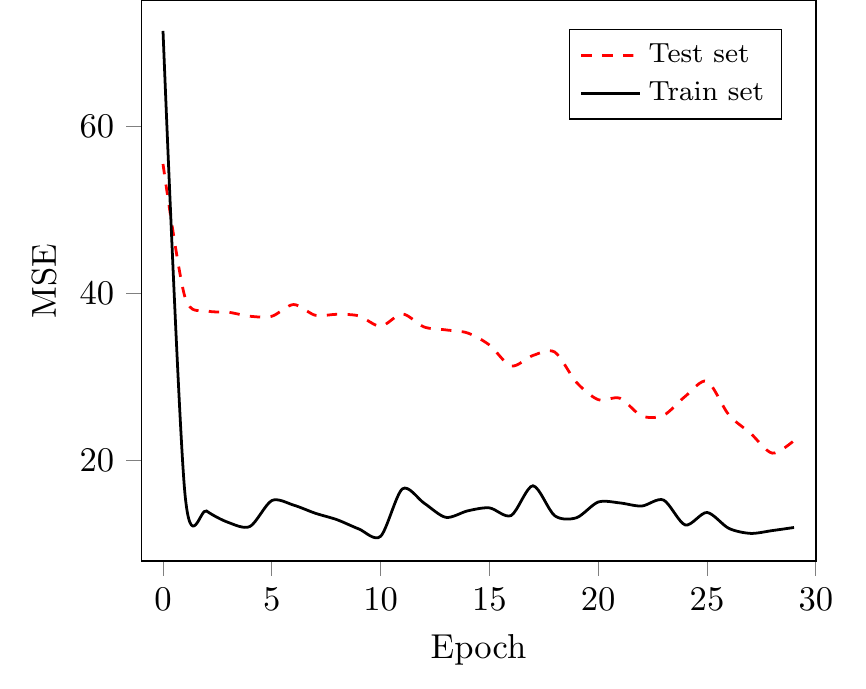}
    \caption{Mean Squared Error during training. 
    The solid line refers to the training set and the dashed one to the test set.}
    \label{fig: MSE}
\end{figure}

\Cref{fig: MSE} shows the \ac{MSE} over 30 training epochs. The results obtained by evaluating the training sets are plotted in dotted lines, while the ones for the \ac{MSE} of the test set are shown in solid lines. 
From both sets of samples, 
% it can be observed that Predictor 2 has a superior performance over Predictor 1. 
% Additionally, 
it is clear from Fig. \ref{fig: MSE} that Predictor has smoother training and converges. 
Also, it can be noticed that with just 8 training epochs, Predictor achieves a satisfactory \ac{MSE}.

\begin{figure}%[h]%t!]
    \centering
    \includegraphics[width=0.9\linewidth]{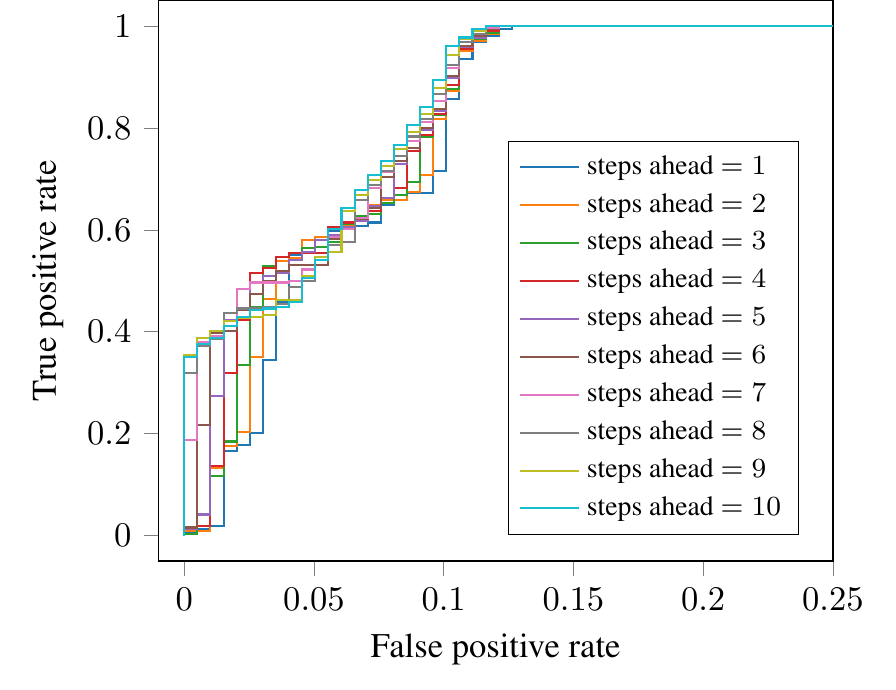}
    \caption{\ac{ROC} curves for a single sub-band and for the different steps ahead. This figure is better viewed in color.}
    \label{fig: ROC}
\end{figure}

Figure \ref{fig: ROC} presents the \ac{ROC} curves of the classifier for the different steps ahead predictions considering only a single sub-band. The classifier performed satisfactorily. Furthermore, it is clear from the Fig. \ref{fig: ROC} that the performance is similar among the presented curves. In other words, sorting copes equally well in predicting channel behaviour within a short term (1 step ahead) or long term (10 steps ahead).

Another run of channel simulation---which is distinct and independent from the first two simulation runs---is provided in an effort to ascertain possible overfitting on the proposed model. The new channel simulation has $4096$ samples, and given the way the model is trained---notice that we purposely do not randomly split the total samples set into two sets, training and testing---an overfitting margin may exist. 
This issue can be evaluated by testing the predictor with new and independent samples.
The obtained results for the the predictor are shown in \Cref{tab: MSE} that presents the mean square error for a specific number of steps ahead. 
In other words, for each of the possible future steps that the proposed model performs a prediction on, the mean square error for the new set of inputs is calculated. 
It can be noticed that the predictor does not show an overfitting behaviour, and has a mean square error that is very close to that obtained for the first set of inputs. 
% 
% For example, the \ac{MSE} for the test set is 10.94 while the one for the new set is 10.6. 
% 
This characteristic is a strong indication that we can train the proposed model in a given environment and use it in an environment that is uncorrelated to the former one. This characteristic needs to be investigated comprehensively.
However, this falls outside the scope of the present work. 
% While we still refer to Fig. \ref{fig: MSE Test}, a slight increase in the \ac{MSE} can be observed as the step ahead increases. For example, in Predictor 2 there is an
% increase of 8.5\% in \ac{MSE} between a step ahead from 1 to 10. The slight deterioration in the prediction's quality is plausible since it is natural for stochastic processes to have a unique and monotonically descending auto-correlation function.
% 
%\vspace{-0.15in}
\begin{table}%[ht]
\renewcommand{\arraystretch}{1.3}
\centering
\caption{Mean Squared Error over the $4096$ Samples from New Channel.}
\label{tab: MSE}
\begin{tabular}{|c|c|}
\hline
\textbf{Step ahead} & \textbf{MSE} \\ \hline
$1$                  & $24.455336$                   \\ \hline
$2$                   & $22.610251$                   \\ \hline
$3$                   & $26.049115$                   \\ \hline
$4$                   & $24.488873$                   \\ \hline
$5$                   & $23.515360$                    \\ \hline
$6$                   & $23.826602$                   \\ \hline
$7$                   & $25.350836$                  \\ \hline
$8$                   & $25.708386$                   \\ \hline
$9$                   & $25.830940$                    \\ \hline
$10$                  & $26.221350$                    \\ \hline
\end{tabular}
\end{table}
\section{Conclusion} \label{sec: Conc}
In this work, we have proposed a simplified way to simulate a mobile radio channel with fading. Moreover, the channel simulation results have revealed typical behaviour of broadband signals under selective fading. Additionally, a predictor and a classifier constituted by layer of \ac{CNN} in sequences have be employed to predict the multi-step fading channel in wireless broadband systems. 
%\vspace{-0.15in}
\bibliographystyle{IEEEtran}
\bibliography{IEEEabrv,References}

\begin{thebibliography}{10}
\providecommand{\url}[1]{#1}
\csname url@samestyle\endcsname
\providecommand{\newblock}{\relax}
\providecommand{\bibinfo}[2]{#2}
\providecommand{\BIBentrySTDinterwordspacing}{\spaceskip=0pt\relax}
\providecommand{\BIBentryALTinterwordstretchfactor}{4}
\providecommand{\BIBentryALTinterwordspacing}{\spaceskip=\fontdimen2\font plus
\BIBentryALTinterwordstretchfactor\fontdimen3\font minus
  \fontdimen4\font\relax}
\providecommand{\BIBforeignlanguage}[2]{{%
\expandafter\ifx\csname l@#1\endcsname\relax
\typeout{** WARNING: IEEEtran.bst: No hyphenation pattern has been}%
\typeout{** loaded for the language `#1'. Using the pattern for}%
\typeout{** the default language instead.}%
\else
\language=\csname l@#1\endcsname
\fi
#2}}
\providecommand{\BIBdecl}{\relax}
\BIBdecl

\bibitem{LiBaolong2021LOOW}
B.~Li, X.~Xue, S.~Feng, and W.~Xu, ``\BIBforeignlanguage{eng}{Layered optical
  {OFDM} with adaptive bias for dimming compatible visible light
  communications},'' \emph{\BIBforeignlanguage{eng}{J. Lightw. Technol.}},
  vol.~39, no.~11, pp. 3434--3444, 2021.

\bibitem{jmoualeu}
J.~M. Moualeu, W.~Hamouda, and F.~Takawira, ``Performance of af relay selection
  with outdated channel estimates in spectrum-sharing systems,'' \emph{IEEE
  Commun. Lett.}, vol.~20, no.~9, pp. 1844--1847, Sep. 2016.

\bibitem{duel-hallen}
A.~Duel-Hallen, ``Fading channel prediction for mobile radio adaptive
  transmission systems,'' \emph{Proc. {IEEE}}, vol.~95, no.~12, pp. 2299--2313,
  Dec. 2007.

\bibitem{Vandenameele895036}
P.~Vandenameele \emph{et~al.}, ``A combined {OFDM/SDMA} approach,''
  \emph{{IEEE} J. Sel. Areas Commun.}, vol.~18, no.~11, pp. 2312--2321, Nov.
  2000.

\bibitem{icwong}
I.~Wong, A.~Forenza, R.~Heath, and B.~Evans, ``Long range channel prediction
  for adaptive {OFDM} systems,'' in \emph{Asilomar Conference on Signals,
  Systems and Computers, 2004.}, vol.~1, 2004, pp. 732--736 vol.1.

\bibitem{Torres2021}
J.~F. Torres, D.~Hadjout, A.~Sebaa, F.~Mart\'{i}nez-\'{A}lvarez, and
  A.~Troncoso, ``Deep learning for time series forecasting: A survey,'' vol.~9,
  no.~1, pp. 3--21, Feb. 2021.

\bibitem{wjiang}
W.~Jiang and H.~D. Schotten, ``Neural network-based fading channel prediction:
  A comprehensive overview,'' \emph{IEEE Access}, vol.~7, pp.
  118\,112--118\,124, 2019.

\bibitem{liulei}
L.~Liu, L.~Cai, L.~Ma, and G.~Qiao, ``Channel state information prediction for
  adaptive underwater acoustic downlink {OFDMA} system: Deep neural networks
  based approach,'' \emph{IEEE Trans. Veh. Technol.}, vol.~70, no.~9, pp.
  9063--9076, Sep. 2021.

\bibitem{yshao}
Y.~Shao, M.-M. Zhao, L.~Li, and M.~Zhao, ``Deep learning based channel
  prediction for {OFDM} systems under double-selective fading channels,'' in
  \emph{2022 Int. Symp. Wireless Commun. Syst. (ISWCS)}, 2022, pp. 1--6.

\bibitem{bultitude20074}
Y.~d.~J. Bultitude and T.~Rautiainen, ``{IST-4-027756 WINNER II D1. 1.2 V1. 2
  WINNER II Channel Models},'' \emph{EBITG, TUI, UOULU, CU/CRC, NOKIA, Tech.
  Rep}, 2007.

\bibitem{Jaeckel6758357}
J.~Stephan \emph{et~al.}, ``Quadriga: A 3-{D} multi-cell channel model with
  time evolution for enabling virtual field trials,'' \emph{{IEEE} Trans.
  Antennas Propag.}, vol.~62, no.~6, pp. 3242--3256, 2014.

\bibitem{Yin7378822}
X.~Yin, C.~Ling, and M.-D. Kim, ``Experimental multipath-cluster
  characteristics of 28-{GHz} propagation channel,'' \emph{{IEEE} Access},
  vol.~3, pp. 3138--3150, 2015.

\bibitem{TSGS2019}
\BIBentryALTinterwordspacing
TSGS, ``{TR 121 915 - V15.0.0 - Digital cellular telecommunications system
  (Phase 2+) (GSM); Universal Mobile Telecommunications System (UMTS); LTE; 5G;
  Release description; Release 15 (3GPP TR 21.915 version 15.0.0 Release
  15)},'' {3GPP}, Tech. Rep., 2019. [Online]. Available:
  \url{https://www.etsi.org/}
\BIBentrySTDinterwordspacing

\end{thebibliography}

\end{document}